\documentclass[preprint,prd,nofootinbib,tightenlines,amsmath]{revtex4}
\usepackage{enumerate}
\usepackage{multirow}

\usepackage{epsfig}
\usepackage{graphics,color}      
\usepackage{verbatim}
\usepackage{amsmath}   
\catcode`\@=11
\def\sla@#1#2#3#4#5{{%
 \setbox\z@\hbox{$\m@th#4#5$}%
 \setbox\tw@\hbox{$\m@th#4#1$}%
 \dimen4\wd\ifdim\wd\z@<\wd\tw@\tw@\else\z@\fi
 \dimen@\ht\tw@
 \advance\dimen@-\dp\tw@ \advance\dimen@-\ht\z@
 \advance\dimen@\dp\z@
 \divide\dimen@\tw@ \advance\dimen@-#3\ht\tw@
 \advance\dimen@-#3\dp\tw@ \dimen@ii#2\wd\z@
 \raise-\dimen@\hbox to\dimen4{%
 \hss\kern\dimen@ii\box\tw@\kern-\dimen@ii\hss}%
 \llap{\hbox to\dimen4{\hss\box\z@\hss}}}}

\def\slashed#1{%
 \expandafter\ifx\csname sla@\string#1\endcsname{\rm ~Re}lax
{\mathpalette{\sla@/00}{#1}}
\fi}
\def\declareslashed#1#2#3#4#5{%
 \expandafter\def\csname sla@\string#5\endcsname{%
#1{\mathpalette{\sla@{#2}{#3}{#4}}{#5}}}}
 \catcode`\@=12
\declareslashed{}{/}{.08}{0}{D}
 \declareslashed{}{/}{.1}{0}{A}
 \declareslashed{}{/}{0}{-.05}{k}
 \declareslashed{}{/}{.1}{0}{\partial}
 \declareslashed{}{\not}{-.6}{0}{f}

\def\lsim{\mathrel {\vcenter {\baselineskip 0pt \kern 0pt
    \hbox{$<$} \kern 0pt \hbox{$\sim$} }}}
\def\gsim{\mathrel {\vcenter {\baselineskip 0pt \kern 0pt
    \hbox{$>$} \kern 0pt \hbox{$\sim$} }}}

\begin{document}

\baselineskip=15pt
\preprint{}

\title{Constraints on new physics from $K \to \pi \nu \bar\nu$}

\author{Xiao-Gang He$^{1,2,3}$\footnote{Electronic address: hexg@phys.ntu.edu.tw},  German Valencia$^{4}$\footnote{Electronic address: german.valencia@monash.edu } and Keith Wong$^{4}$}

\affiliation{$^1$ 
Department of Physics, National Taiwan University, Taipei 10617}

\affiliation{$^2$
Physics Division, National Center for Theoretical Sciences, Hsinchu 30013}

\affiliation{$^3$
Tsung-Dao Lee Institute, and School of Physics and Astronomy, Shanghai Jiao Tong University, Shanghai 200240}

\affiliation{$^4$ School of Physics and Astronomy, Monash University, Melbourne VIC-3800 }

\date{\today}

\vskip 1cm
\begin{abstract}

We study generic effects of new physics on the rare decay modes $K_L \to \pi^0 \nu \bar\nu$ and $K^+ \to \pi^+ \nu \bar\nu$. We discuss several cases:  left-handed neutrino couplings; right handed neutrino couplings; neutrino lepton flavour violating (LFV) interactions; and $\Delta I =3/2$ interactions. The first of these cases has been studied before as it covers many new physics extensions of the standard model; the second one requires the existence of a new light (sterile) right-handed neutrino and  its contribution to both branching ratios is always additive to the SM. The case of neutrino LFV couplings introduces a CP conserving contribution to $K_L \to \pi^0 \nu \bar\nu$ which affects the rates in a similar manner as a right handed neutrino as neither one of these interferes with the standard model amplitudes. Finally, we consider new physics with $\Delta I =3/2$ interactions to go beyond the Grossman-Nir bound. We find that the rare kaon rates are only sensitive to new physics scales up to a few GeV for this scenario.

\end{abstract}

\pacs{PACS numbers: }

\maketitle
    

\newpage

\section{Introduction}

In the standard model (SM), the rare decay modes $K \to \pi \nu\overline\nu$ proceed dominantly via a short distance contribution from a  top-quark intermediate loop.
This allows a precise calculation of the rates in terms of SM parameters  \cite{Hagelin:1989wt,Littenberg:1989ix}.  The effective Hamiltonian responsible for these transitions in the SM is frequently written as
\begin{eqnarray}
{\cal H}=\frac{G_F}{\sqrt{2}}\frac{2\alpha}{\pi s^2_W}V^\star_{ts}V_{td} X(x_t)\bar{s}\gamma_\mu P_L d \sum_\ell \bar{\nu}_\ell\gamma^\mu P_L \nu_\ell.
\label{smH}
\end{eqnarray}
It follows that the branching ratios can then be written as (we use the notation ${\cal B}_{K^+}={\cal B}\left( K^+\rightarrow\pi^+\nu\overline\nu(\gamma)\right) $ and ${\cal B}_{K_L}={\cal B}(K_L \to \pi^0 \nu \overline\nu)$ throughout this paper),
\begin{eqnarray}
{\cal B}_{K^+} &=&\tilde\kappa_+\left[\left(\frac{{{\rm Im}(V^{\star}_{ts}V_{td}X_t)}}{{\lambda^5}}\right)^2
+\left(\frac{{{\rm Re}(V^{\star}_{cs}V_{cd})}}{{\lambda}}P_c+\frac{{{\rm Re}(V^{\star}_{ts}V_{td}X_t)}}{{\lambda^5}}\right)^2\right], \nonumber \\
{\cal B}_{K_L}&=&  \kappa_L \left(\frac{{{\rm Im}(V^{\star}_{ts}V_{td}X_t)}}{{\lambda^5}}\right)^2.
\label{smrates}
\end{eqnarray}
In these equations, the hadronic matrix element of the quark current is written in terms of the well measured semileptonic $K_{e3}$ rate  and is part of the overall constants $\tilde\kappa_+$ and $\kappa_L$. Modern calculations of the parameters in these equations result in: $\tilde\kappa_+ = 0.517 \times 10^{-10}$ which includes long distance QED corrections \cite{Mescia:2007kn}, and $\kappa_L=2.23 \times 10^{-10}$; the Inami-Lim function for the short distance top-quark contribution \cite{Inami:1980fz} including NLO QCD corrections \cite{Buchalla:1998ba} and the  two-loop electroweak correction \cite{Brod:2010hi}, result in $X_t=1.48$; and all known effects of the charm-quark contributions \cite{Buras:2006gb,Brod:2008ss,Isidori:2005xm,Falk:2000nm} in 
$P_c = 0.404$. Finally, $\lambda\approx 0.225$ is the usual Wolfenstein parameter.\footnote{Uncertainties for these quantities can be found in the references.}

Our estimate for these branching ratios within the SM, using the latest CKMfitter input \cite{Charles:2011va}, is 
\begin{eqnarray}
{\cal B}_{K^+} &=&(8.3 \pm 0.4) \times 10^{-11},
\nonumber \\
{\cal B}_{K_L}&=&(2.9\pm 0.2)\times 10^{-11}.
\end{eqnarray}
These numbers are to be compared with the current experimental results for the charged \cite{Adler:2000by,Adler:2001xv,Anisimovsky:2004hr,Artamonov:2009sz} (measured by BNL 787 and BNL 949) and neutral  \cite{Ahn:2009gb} modes (from KEK E391a),
\begin{eqnarray}
{\cal B}_{K^+}&=& (1.73^{+1.15}_{-1.05})\times 10^{-10},
\nonumber \\
{\cal B}_{K_L}&\leq& 2.6 \times 10^{-8} {\rm ~at~}90\%{\rm ~c.l.}
\end{eqnarray}
An interesting correlation between these two modes was pointed out by Grossman and Nir (GN), namely that ${\cal B}_{K_L} \lsim 4.4 \ {\cal B}_{K^+}$ which is satisfied in a nearly model independent way \cite{Grossman:1997sk}. \footnote{It was recently noted that the GN bound applied to the experimental result for $K^+\rightarrow\pi^+\nu\overline\nu$ needs to treat a possible two body intermediate state separately \cite{Fuyuto:2014cya}.} 

In this paper we revisit these modes in the context of generic new physics motivated by the new results that are expected soon for the charged mode from NA62 at CERN and for the neutral mode from KOTO in Japan. Our paper is organised in terms of the neutrino interactions as follows: in section~II we briefly review extensions of the SM in which the neutrino interactions are left handed and flavour conserving; in section~III we consider extensions of the SM with right-handed neutrino interactions; in section~IV we discuss the lepton flavour violating case. In section~V we study interactions that violate the GN bound and finally, in section~VI, we conclude.

\section{New physics with lepton flavour conserving left-handed neutrinos}

\begin{figure}[!htb]
\begin{center}
\includegraphics[width=6cm]{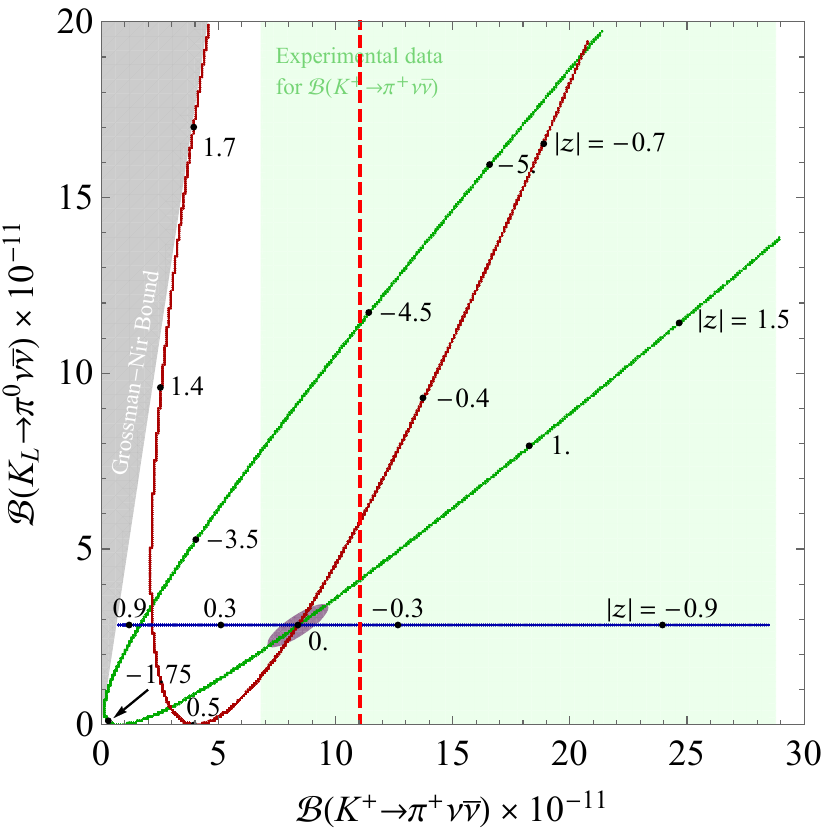}
\end{center}
\caption{New physics with lepton flavour conserving left-handed neutrinos. The green line illustrates the case $X_N$ real, the red line corresponds to $X_N$ having a phase equal to that of the $\lambda_t$ (central value) and the blue line to $X_N$ having a phase equal to minus that of the $\lambda_t$. For comparison the purple marks the SM $1\sigma$ region and the green marks the 90\% c.l. from BNL-787 combined with BNL-949.  Finally the vertical dashed red line marks a possible future limit for ${\cal B}_{K^+}$ at 1.3 times the SM.}
\label{f:case1}
\end{figure}

In this case the effective Hamiltonian describing the effects of the new physics (NP) takes the form
\begin{eqnarray}
{\cal H}_{eff}=\frac{G_F}{\sqrt{2}}\frac{2\alpha}{\pi s^2_W}V^\star_{ts}V_{td} X_N \bar{s}\gamma_\mu  d \sum_\ell \bar{\nu}_\ell\gamma^\mu P_L \nu_\ell,
\label{npcase1}
\end{eqnarray}
where the parameters encoding the NP are collected in $X_N$ and the overall constants have been chosen for convenience. Notice that this form is valid for both left-handed and right-handed quark currents as only the vector current is operative for the $K\to \pi$ transition. Numerically it is then possible to obtain the rates from the SM result, Eq.~\ref{smrates}, via the substitution $X(x_t)\to X(x_t) + X_N$. This has been done in the literature for a variety of models \cite{Buras:2015yca} so we will not dwell on this case here. In Figure~\ref{f:case1} we illustrate the results. In general $X_N\equiv ze^{i\phi}$ and the parameterisation in Eq.~\ref{npcase1} implies that $\phi=0$ corresponds to NP with the same phase as $\lambda_t=V^\star_{ts}V_{td}$. The green curve corresponds to $\phi=0$ (so called MFV in \cite{Buras:2015yca}) and its two branches correspond to constructive and destructive interference with the charm-quark contribution in Eq.~\ref{smrates}. The tick marks on the curve mark  values of $|X_N|=z$. If we allow for an arbitrary phase, this type of NP can populate the entire area below the GN bound, making it nearly impossible to translate a non-SM measurement into values of $z$ and $\phi$.

We illustrate two more situations: the blue line shows $\phi$ being {\it minus} the phase of $\lambda_t$, which corresponds to CP conserving NP which does not contribute to the neutral kaon mode. The red line shows $\phi$ being the same as the phase of $\lambda_t$, which corresponds to NP which doubles the SM phase. Interestingly this case nearly saturates the GN bound.  For comparison, we show the purple oval representing the $1\sigma$ SM allowed region as predicted using the parameters and uncertainties in CKMfitter \cite{Charles:2011va}. For the NP, however, we have only included the SM central values in Eq.~\ref{npcase1}. Allowing the SM parameters to vary in the rates that include NP, turns the green line into an arc-shaped region as can be seen in Ref.~ \cite{Buras:2015yca} for example. 

Finally we have included in the plot a vertical red dashed line which marks a 30\% uncertainty from the SM central value. This number has been chosen as it corresponds to the statistical uncertainty that can be achieved with 10 events that agree with the SM, in the ball park of what is expected from NA62. 

\section{A light right handed neutrino}

\begin{figure}[!htb]
\begin{center}
\includegraphics[width=6cm]{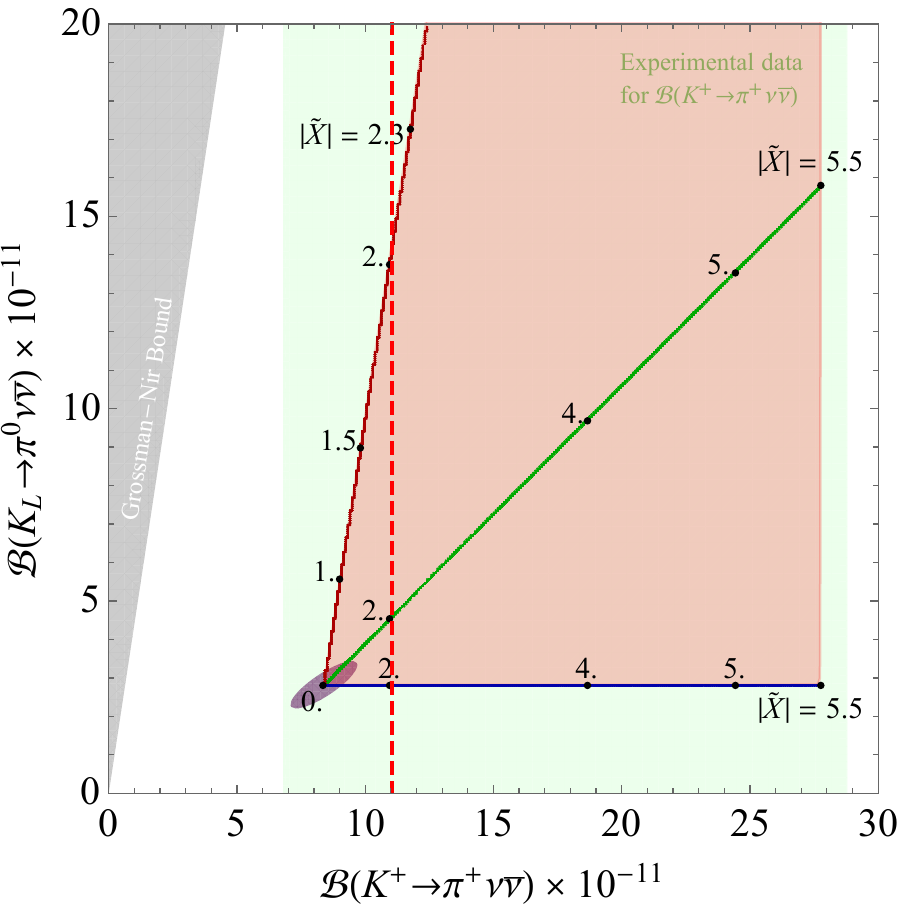}
\end{center}
\caption{New physics with one light right-handed neutrino. The green line illustrates the case $\tilde X$ real and the pink region illustrates the case $|\tilde X|\leq 5.5$.  The purple marks the SM $1\sigma$ region and the green marks the 90\% c.l. from BNL-787 combined with BNL-949. The red and blue lines on the boundary of the pink region correspond to a new physics phase given by $\phi+\phi_{\lambda_t}=(\pi/2{\rm ~or~}3\pi/2)$ and $\phi+\phi_{\lambda_t}=(0{\rm ~or~}\pi)$ respectively. Finally the vertical dashed red line marks a possible future limit for ${\cal B}_{K^+}$ at 1.3 times the SM.}
\label{f:case2}
\end{figure}

In models which contain a light right handed neutrino the effective Hamiltonian can be written as 
\begin{eqnarray}
{\cal H}_{eff}=\frac{G_F}{\sqrt{2}}\frac{2\alpha}{\pi s^2_W}V^\star_{ts}V_{td} \frac{1}{2}\bar{s}\gamma_\mu  d\left( X_t \sum_\ell \bar{\nu}_\ell\gamma^\mu P_L \nu_\ell+ \tilde{X} \bar{\nu}_R\gamma^\mu P_R \nu_R\right),
\label{npcase2}
\end{eqnarray}
where the first term is the SM, the new physics is parameterised by $\tilde X$ and its coupling to quarks can be through either a left or right handed current. In writing Eq.~\ref{npcase2} we have assumed that there is only one new neutrino and that its mass is negligible. The rates for the rare kaon decay modes follow immediately,
\begin{eqnarray}
{\cal B}_{K^+}(\nu_{RH}) &=&{\cal B}_{K^+}  (SM)+\frac{\tilde\kappa_+}{3}\left|\frac{\lambda_t\tilde{X}}{\lambda^5}\right|^2\nonumber\\
{\cal B}_{K_L}(\nu_{RH})&=& {\cal B}_{K_L} (SM) + \frac{ \kappa_L}{3} \left(\frac{{\rm ~Im}\lambda_t\tilde{X}}{\lambda^5}\right)^2 \label{ratescase2}
\end{eqnarray}
where the $1/3$ accounts for the fact that we have only one right handed light neutrino (a factor of 3 from summing over the left-handed neutrinos is hiding in  $\tilde\kappa_+$ and $\kappa_L$). In the result, Eq.~\ref{ratescase2}, we see that this type of NP can only increase the rates, as it does not interfere with the SM, and this is illustrated in Figure~\ref{f:case2}. As in the previous case, we have chosen a parameterisation in Eq.~\ref{npcase2} in which $\tilde{X}\equiv |\tilde{X}|e^{i\phi}$ and $\phi=0$ corresponds to the NP having the same phase as $\lambda_t$. The green line in the figure corresponds to $\phi=0$ and the tick marks show that a maximum value of $|\tilde{X}|\lsim 5.5$ is allowed by the current BNL 90\% c.l. limit on the charged rate, and that this number can be reduced to $|\tilde{X}|\lsim 2$ with about ten events. The pink region covers the parameter space $|\tilde X| \leq 5.5$ with an arbitrary phase, and we show two more lines near the boundary of this region. The red line is obtained for $\phi+\phi_{\lambda_t}=(\pi/2{\rm ~or~}3\pi/2)$; 
 whereas the blue line occurs for $\phi+\phi_{\lambda_t}=(0{\rm ~or~}\pi)$, for which there is no new contribution to the neutral mode.

Within the specific model detailed in the Appendix, the effect of the additional neutrino contributes both via a flavour changing tree-level $Z^\prime$ exchange and a one-loop $Z^\prime$ penguin and can be written as,
\begin{eqnarray}
\tilde{X}=-\left(\frac{M_Z^2}{M_{Z^\prime}^2}\cot^2\theta_R\right)\left(\frac{s^2_W}{2}I(\lambda_t,\lambda_H)+\frac{\pi s^4_W}{\alpha}
\frac{V^{d\star}_{Rbs}V^d_{Rbd}}{V^\star_{ts}V_{td}}\right).
\label{ourxt}
\end{eqnarray}
The overall strength of the $Z^\prime$ coupling is parameterised by $\cot\theta_R\lsim 20$, where the upper limit arises from requiring the interaction to remain perturbative \cite{He:2002ha}. This, combined with the CMS limit on a $Z^\prime$ that decays to tau-pairs $M_{Z^\prime}\gsim 1.7$~TeV \cite{Khachatryan:2016qkc}, implies that the factor in the first bracket of Eq.~\ref{ourxt} can be of order one.  The tree-level contribution (second term in the second bracket) is constrained to be small by $B_s$-mixing and $B_d$-mixing, $|V^{d\star}_{Rbs}V^d_{Rbd}/(V^\star_{ts}V_{td})|\lsim 3\times 10^{-3}$  \cite{He:2006bk}. The Inami-Lim factor appearing in the $Z^\prime$ penguin, $I(\lambda_t,\lambda_H)$, is less constrained and can be of order 10 \cite{He:2004it}. All in all, in our model
the magnitude of $\tilde{X}$ can be order one but its phase is limited by the size of the tree contribution. This provides an example of NP in which a measurement of the two rates can be mapped to parameters in the model.

The existence of an additional light neutrino can, in general, have other observable consequences. As we show in Ref.~\cite{He:2012zp},  the invisible $Z$ width constrains the mixing between the $Z$ and $Z^\prime$ bosons in our model. This mixing, however, does not alter the leading contributions to $\tilde X$ shown in Eq.~\ref{ourxt}. In essence the $Z$ width does not constrain this additional light neutrino because it is sterile as far as the SM interactions are concerned. 
A new  light right-handed neutrino also contributes to the effective number of neutrino species $\Delta N_{eff}$ which is  constrained by  cosmological considerations. In Ref.~\cite{He:2017bft} we show that this constraint can also be evaded if the new neutrino mixes dominantly with the tau-neutrino and not with the muon or electron neutrinos.

\section{Neutrino lepton flavour violating interactions}

\begin{figure}[!htb]
\begin{center}
\includegraphics[width=6cm]{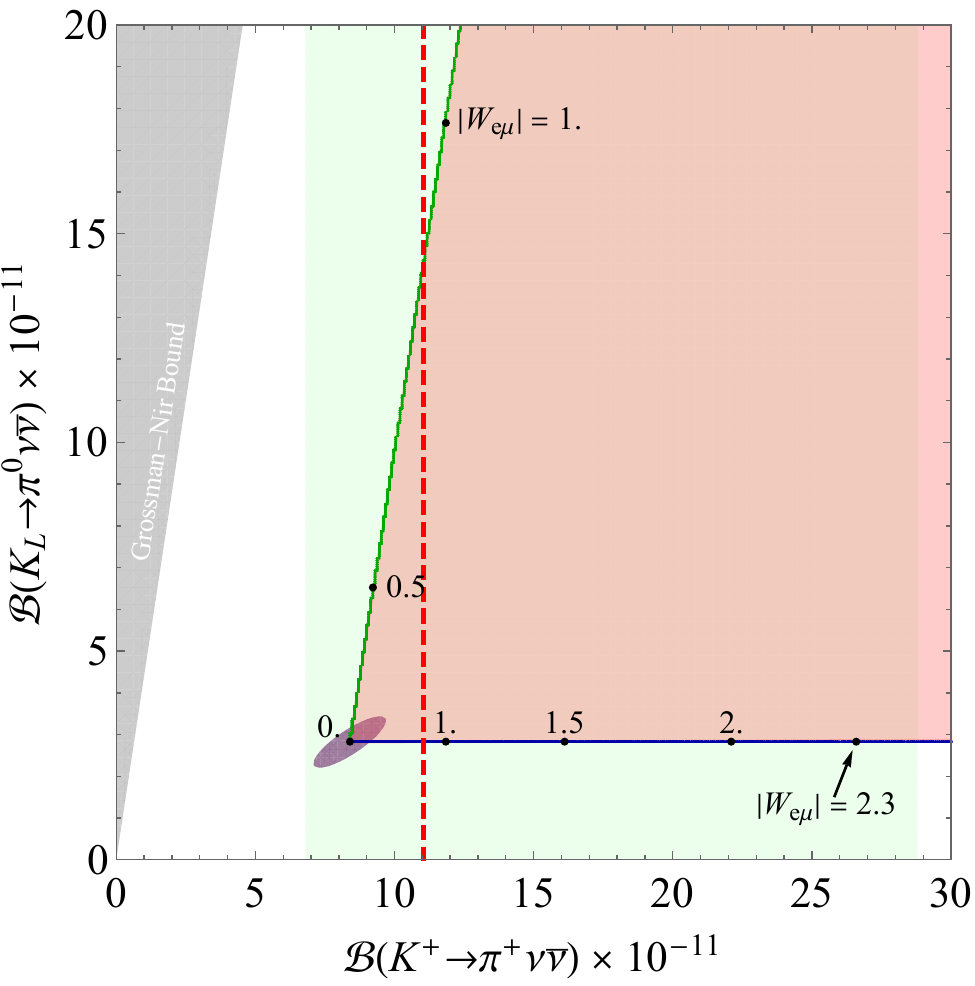}
\end{center}
\caption{New physics with lepton flavour violation. The pink shaded region is allowed for $W_{e\mu,\mu e}$  satisfying $|W_{e\mu,\mu e}| \leq 6$ with arbitrary phases. The left boundary of the region (green line) corresponds to the case $W_{e\mu}=-W_{\mu e}^\star$, whereas the blue boundary (horizontal line) occurs for $W_{e\mu}=W_{\mu e}^\star$. As before, the purple marks the SM $1\sigma$ region, the green marks the 90\% c.l. from BNL-787 combined with BNL-949 and the vertical dashed red line illustrates a possible future limit for ${\cal B}_{K^+}$ at 1.3 times the SM.}
\label{f:case3}
\end{figure}

Another possibility consists of interactions that violate lepton flavour conservation in the neutrino sector. These are particularly interesting because they can yield CP conserving contributions to the  $K_L\to\pi^0\nu\bar\nu$ decay. In this case it is convenient to  write 
\begin{eqnarray}
{\cal H}_{eff}=\frac{G_F}{\sqrt{2}}\frac{2\alpha}{\pi s^2_W} \frac{1}{2}\bar{s}\gamma_\mu  d\left( \sum_\ell \left(V^\star_{ts}V_{td}X_t +\lambda^5 W_{\ell\ell}\right)\bar{\nu}_\ell\gamma^\mu P_L \nu_\ell+ 
\lambda^5\sum_{i\neq j} W_{ij} \bar{\nu}_i\gamma^\mu P_L \nu_j \right) + {\rm ~h.~c.}
\label{npcase3}
\end{eqnarray}
to normalise the strength of the NP to that of the SM but without inserting the SM phase into the new couplings. This then results in
\begin{eqnarray}
&&{\cal B}_{K^+}({LFV}) ={\cal B}_{K^+} (SM)+\frac{\tilde\kappa_+}{3}\sum_{i\neq j}\left|W_{ij}\right|^2  \nonumber \\
&&{\cal B}_{K_L}(LFV)= {\cal B}_{K_L}(SM) + \frac{\kappa_L}{3}\sum_{i\neq j}\left|\frac{(W_{ij}-W^\star_{ji})}{2}\right|^2
\label{ratescase3}
\end{eqnarray}
where  again a factor of $1/3$ compensates for the factor of 3 hiding in $\tilde\kappa_+$ and $\kappa_L$. These lepton flavor violating contributions (proportional to $W_{ij}$, $i\neq j$) produce a very similar pattern of corrections as the case of the right handed neutrino Eq.~\ref{npcase2}. This LFV contribution to the neutral mode is maximised when
\begin{eqnarray}
W_{ij}=-W_{ji}^\star,
\end{eqnarray}
and we illustrate this scenario in Figure~\ref{f:case3}.  
The green line corresponds to the case $W_{e\mu}=-W_{\mu e}^\star$ and the dots mark values of $|W_{e\mu}|$. The allowed region when only $W_{e\mu,\mu e}$ is allowed to be non-zero and satisfying $|W_{e\mu,\mu e}| \leq 6$ with arbitrary phases is shown in pink. The blue line, where the neutral kaon rate is unaffected, occurs for $W_{ij}=W_{ji}^\star$.

Neither the LFV nor the right-handed neutrino scenarios interferes with the SM amplitude, so they both result in additive corrections to the rates. 
We can illustrate the correspondence between the two cases  by considering the red line of Figure~\ref{f:case2} for which the phase of $\tilde{X}$ plus the phase of $\lambda_t$ equals $\pi/2,3\pi/2$ and therefore ${\rm Re}(\lambda_t\tilde{X})=0$. This line matches the green line of Figure~\ref{f:case3} for $W_{e\mu}=-W_{\mu e}^\star$, and $|W_{e\mu}|\sim 1$ is equivalent to $|\tilde{X}|\sim 2.3$.

Figure~\ref{f:case3} indicates that this model can have important effects for $W_{ij}\sim {\cal O}(1)$. In terms of the leptoquark couplings shown in the Appendix, $c_{ij}$ is of order
\begin{eqnarray}
c_{ij} \sim \frac{G_F}{\sqrt{2}}\frac{2\alpha}{\pi s^2_W}V^\star_{ts}V_{td} W_{ij}\sim \frac{g^2 }{(83.5~{\rm TeV})^2}
\end{eqnarray}
implying that for leptoquark couplings of electroweak strength, these rare kaon modes are sensitive to leptoquark masses up to about 80~TeV.

\section{Beyond the Grossman-Nir bound}

The hadronic transition between a kaon and a pion can be mediated in general by an operator that changes isospin by $1/2$ or by $3/2$. The ratio of matrix elements follows from the Clebsch-Gordan coefficients
\begin{eqnarray}
\frac{<\pi^0|{\cal O}_{\Delta I = 1/2}|K^0>}{<\pi^+|{\cal O}_{\Delta I = 1/2}|K^+>} = -\frac{1}{\sqrt{2}}, &&
\frac{<\pi^0|{\cal O}_{\Delta I = 3/2}|K^0>}{<\pi^+|{\cal O}_{\Delta I = 3/2}|K^+>}= \sqrt{2}
\end{eqnarray}
and the GN bound follows from the first of these equations, appropriate for the $\bar{s}d$ isospin structure of dimension six effective Hamiltonians of the cases discussed so far. Long distance contributions in the SM can violate this isospin relation but they are known to be small~\cite{Lu:1994ww}.  Long distance contributions within the SM can also produce CP conserving contributions to $K_L \to \pi^0 \nu \overline\nu$ due to different CP properties of the relevant operators but these effects are also known to be small \cite{Buchalla:1998ux}. 

When the $K\to \pi$ transition is mediated by a vector current, as in the short distance SM of Eq.\ref{smH}, the $K_L\to \pi^0\nu\bar\nu$ decay is CP violating due to the CP transformation properties of the current: $\bar{s}\gamma_\mu d \overset{CP}{\longleftrightarrow} -\bar{d}\gamma^\mu s$. In the same manner $K_L\to \pi^0\nu\bar\nu$ is CP conserving when mediated by a scalar density as $\bar{s}d \overset{CP}{\longleftrightarrow} \bar{d} s$ \cite{Kiyo:1998aw}.\footnote{The operator discussed in this reference, $\bar s_R d_L\bar \nu_R\nu_L$, can be generated by leptoquark exchange at tree level in models which also have right handed neutrinos. Its effects satisfy the GN bound and, as it does not interfere with the SM, produces changes to the rates similar to the ones already discussed for LFV interactions.} These CP conserving contributions do not interfere with the SM amplitudes and cannot violate GN \cite{He:2020jzn}.

\begin{figure}[!htb]
\begin{center}
\includegraphics[width=6cm]{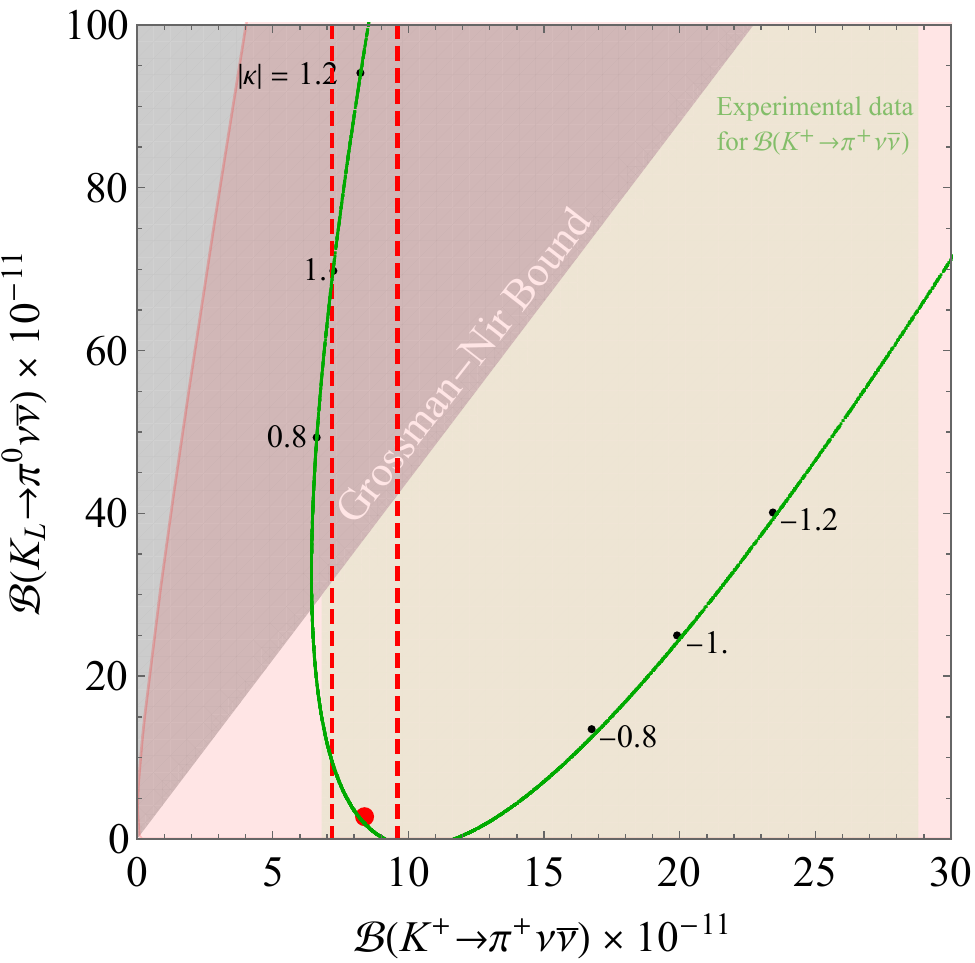}
\end{center}
\caption{Rates covered by Eq.~\ref{npcase4} are illustrated in pink along with the BNL result in green and the GN exclusion in grey. The SM central values are shown as the large red dot and the dashed vertical lines correspond to $\pm 3\sigma$ from the central SM value of ${\cal B}_{K^+}$. The green curve for  $\phi_\kappa=45^\circ$ is chosen to illustrate values that can produce ${\cal B}_{K_L}\sim 10^{-9}$ while keeping  ${\cal B}_{K^+}$ near its SM value.}
\label{f:bgn}
\end{figure}

To construct a $\Delta S=1, \Delta I =3/2$ operator one needs at least four quarks, and we give an example that leads to a CP violating $K_L\to \pi^0\nu\bar\nu$ in the appendix. Many operators with these properties can occur beyond the SM  and are discussed at length in \cite{He:2020jzn,He:2020jly}.  
The effect of this example on the $K\to \pi \nu \bar\nu$ modes can be written as (when added to the SM)
\begin{eqnarray}
&&{\cal B}_{K^+}\ =\ \tilde\kappa_+\left[\left(\frac{{\rm Im}(V^{\star}_{ts}V_{td}X_t)}{\lambda^5}+{\rm Im}\ \kappa^{\frac{3}{2}}\right)^2 
 +\left(\frac{{\rm Re}(V^{\star}_{cs}V_{cd})}{\lambda}P_c+\frac{{\rm Re}(V^{\star}_{ts}V_{td}X_t)}{\lambda^5} 
+{\rm Re}\ \kappa^{\frac{3}{2}}
\right)^2\right],  \nonumber \\
&&{\cal B}_{K_L}\ =\  \kappa_L\left[ \left(\frac{{\rm Im}(V^{\star}_{ts}V_{td}X_t)}{\lambda^5}+2\ {\rm Im}\ \kappa^{\frac{3}{2}}\right)^2\right].
\label{npcase4}
\end{eqnarray}
These relations are illustrated in Figure~\ref{f:bgn} where the range  covered by the rates of Eq.~\ref{npcase4} is shown in pink along with the BNL result in green and the GN exclusion in grey. The SM central values are shown as the large red dot (the one sigma SM region is small on the scale of this plot) and the dashed vertical lines correspond to $\pm 3\sigma$ from the central SM value of ${\cal B}_{K^+}$. The green curve for  $\phi_\kappa=45^\circ$ is chosen to illustrate values that can produce ${\cal B}_{K_L}\sim 10^{-9}$ while keeping  ${\cal B}_{K^+}$ near its SM value.
When $\Delta I=3/2$ interactions are present, the GN bound is no longer valid. In addition, with the pattern of NP appearing in Eq.~\ref{npcase4} and illustrated in Figure~\ref{f:bgn}, it is possible to keep the charged rate close to the SM while making the neutral rate as large as desired. 
Considering the dimension eight operator of the appendix, Eq.~\ref{leff27}, the NP coupling reads,
\begin{eqnarray}
\kappa^{\frac{3}{2}} &=& \frac{g_{NP}}{\Lambda^4}f_\pi f_K m_K^2 r_{ps}\frac{\sqrt{2}\pi s^4_W}{\alpha \lambda^5}.
\end{eqnarray}
Figure~\ref{f:bgn} shows that the rare kaon rates are sensitive to  $\kappa^{\frac{3}{2}}\sim 1$. With $g_{NP}\sim 1$ this then implies they are sensitive to  a NP scale of order $\Lambda \sim 2.3$~GeV. Given that this scale is only a few times larger than $\Lambda_{QCD}$, our result is the same for the different types of possibilities discussed in the appendix, and it shows that even though this scenario is possible in principle, its effects are extremely small. The conclusion is that a violation of the GN bound is completely implausible without a new few GeV  particle that carries isospin.

\section{Conclusions}

We have considered how different types of new physics can affect the rates of the rare kaon decay modes $K\to \pi \nu\bar\nu$. Our findings can be summarised as follows.

\begin{itemize}
\item Lepton flavour conserving left-handed neutrinos. This case allows interference between the NP and the SM and can produce values for the rates anywhere below the GN bound as shown in Figure~\ref{f:case1}. Measurement of these rates can result in clear evidence for NP but an interpretation of the results in terms of NP parameters will be much harder.

\item A light right-handed neutrino. In this case the NP cannot interfere with the SM so the resulting rates are always larger than the SM values. A measurement of the charged mode by NA62 
which agrees with the SM with roughly ten events would place strong new constraints on the magnitude of the RH neutrino interactions. It would also result in an upper bound for the neutral rate ${\cal B}_{K_L} \lsim14 \times 10^{-11}$.

\item Neutrino lepton flavour violating interactions. These scenarios are very similar to new right handed neutrinos as they also do not interfere with the SM. A measurement of the charged mode by NA62 would thus produce equivalent constraints as in the case with RH neutrinos. Models that generate this type of interactions, such as the LQ discussed in the appendix, are likely to also generate flavour conserving left-handed neutrino interactions. In that case there is no clear connection between a measurement and NP parameters.

\item We found interactions with $\Delta I =3/2$ transitions that can violate the GN bound. This scenario would dilute the correlation between charged and neutral modes requiring a direct measurement of the neutral mode  to constrain it. Our study in terms of effective operators suggests  that only very low values of the new physics scale, of order a few GeV, would be observable.

\end{itemize}

\begin{acknowledgments}

This work was supported in part by the ResearchFirst undergraduate research program at Monash University and by the Australian Research Council. 
X.G.H. was supported in part by the MOST (Grant No. MOST104-2112-M-002-015-MY3 and 106-2112-M-002-003-MY3 ), and in part by Key Laboratory for Particle Physics,
Astrophysics and Cosmology, Ministry of Education, and Shanghai Key Laboratory for Particle
Physics and Cosmology (Grant No. 15DZ2272100), and in part by the NSFC (Grant Nos. 11575111 and 11735010).

\end{acknowledgments}

\appendix

\section{A model with a right handed neutrino}

The model has been described in detail elsewhere \cite{He:2002ha,He:2003qv}, here we summarise its salient features. The gauge group  is $SU(3)_C\times SU(2)_L\times SU(2)_R \times U(1)_{B-L}$, but the three generations of fermions are chosen to transform differently to single out the third generation. In the weak interaction basis, the first two generations of quarks $Q_L^{1,2}$, $U_R^{1,2}$, $D_R^{1,2}$ transform as $(3,2,1)(1/3)$, $(3,1,1)(4/3)$ and $(3,1,1)(-2/3)$, and the leptons $L_L^{1,2}$, $E_R^{1,2}$ transform as $(1,2,1)(-1)$ and $(1,1,1)(-2)$. The third generation, on the other hand, transforms as  $Q_L^3\;(3,2,1)(1/3)$, $Q^3_R\;(3,1,2)(1/3)$, $L^3_L\;(1,2,1)(-1)$ and $L^3_R\;(1,1,2)(-1)$. In this way $SU(2)_R$  acts only on the third generation. 

To separate the symmetry breaking scales of $SU(2)_L$ and $SU(2)_R$, there are two 
Higgs multiplets $H_L\; (1,2,1)(-1)$ and $H_R\;(1,1,2)(-1)$ with respective vevs $v_L$ and $v_R$. An additional bi-doublet $\phi\;(1,2,2)(0)$ scalar with vevs $v_{1,2}$ is  needed to provide mass to the fermions. Since both $v_1$ and $v_2$ are required to be non-zero for fermion mass generation, the  $W_L$ and $W_R$ gauge bosons of $S(2)_L$ and $SU(2)_R$ will mix with each other. In terms of the mass eigenstates $W$ and $W'$, the mixing can be parameterized as
\begin{eqnarray}
W_L &=& \cos\xi_W W - \sin \xi_W W'\;,\nonumber\\
W_R &=& \sin\xi_W W +\cos\xi_W W'\;.
\end{eqnarray}
In the mass eigenstate basis the quark-gauge-boson interactions are 
given by,
\begin{eqnarray}
{\cal L}_W&=& - \frac{g_L}{ \sqrt{2}} \bar U_L \gamma^\mu V_{KM} D_L
(\cos\xi_W W^{+}_\mu - \sin\xi_W W^{'+}_\mu)\nonumber\\
&&-\frac{g_R}{ \sqrt{2}}
\bar U_{R} \gamma^\mu V_{R} D_{R}
(\sin\xi_W W^{+}_\mu + \cos\xi_W W^{'+}_{\mu}) ~+~{\rm h.~c.},\nonumber \\
{\cal L}_Z&=& \frac{g_L}{ 2} \tan\theta_W (\tan\theta_R + \cot\theta_R) (\sin\xi_Z Z_\mu + \cos\xi_Z Z^\prime_\mu)\nonumber\\
&&\times (\bar d_{R_i} V^{d*}_{Rbi} V^d_{Rbj}\gamma^\mu d_{R_j} - \bar u_{R_i} V^{u*}_{Rti} V^u_{Rtj}\gamma^\mu u_{R_j} 
)\;,
\label{cccoup}
\end{eqnarray}
where $U = (u,\;\;c,\;\;t)$, $D = (d,\;\;s,\;\;b)$, $V_{KM}$ is
the Kobayashi-Maskawa mixing matrix and $V_R \equiv (V_{Rij})=(V^{u*}_{Rti}V^{d}_{Rbj})$ with $V^{u,d}_{Rij}$ the unitary matrices
which rotate the right handed quarks $u_{Ri}$ and $d_{Ri}$ from the weak  to the mass eigenstate basis. 

The model has  three left-handed neutrinos $\nu_{L_i}$ and one right-handed neutrino $\nu_{R_3}$.
Additional scalars $\Delta_L\;(1,3,1)(2)$ and $\Delta_R\;(1,1,3)(2)$ with vevs $v^{L,R}_\Delta$ are needed to generate neutrino masses. In order for this model to contribute to the rare kaon decay modes discussed here, we  need the right-handed neutrino 
to be light and thus requires $v^{L,R}_\Delta$ to be small. The mass eigenstates $(\nu^m_L, (\nu^m_{R_3})^c)$  
 are related by a unitary transformation to the weak eigenstates as
\begin{eqnarray}
\left (\begin{array}{c}
\nu_L\\
\nu^c_{R_3}
\end{array}
\right )
= \left ( \begin{array}{cc}
U_L&U_{RL}\\
U_{LR}&U_R
\end{array}
\right )
\left (\begin{array}{c}
\nu^m_L\\
(\nu^m_{R_3})^c
\end{array}
\right )\;.
\end{eqnarray}
In our model $U_L= (U_{Lij})$, $U_{RL} = (U_{RLi3})$ and $U_{LR} = (U_{LR3i})$ and $U_R = (U_{R33})$ are
$3\times 3$, $3\times 1$, $1\times 3$ and $1\times 1$ matrices, respectively.

Rotating the charged leptons from their weak eigenstates $\ell_{L,R}$ to their mass eigenstates $\ell^m_{L,R}$, with $\ell_{L,R} = V^\ell_{L,R} \ell^m_{L,R}$, the lepton interaction with $W$ and $W'$ becomes
\begin{eqnarray}
{\cal L}_W &=& - \frac{g_L}{ \sqrt{2}} (\bar \nu_L \gamma^\mu U^{\ell \dagger}\ell_L 
+ \bar \nu_{R3}^c \gamma^\mu U^{\ell *}_{RLj3}\ell_{Lj})
(\cos\xi_W W^{+}_\mu - \sin\xi_W W^{'+}_\mu) \nonumber \\
&-&\frac{g_R}{ \sqrt{2}}
(\bar \nu^c_{Li} \gamma^\mu  U^\ell_{LRij} \ell_{Rj} + \bar \nu_{R3} \gamma^\mu U^\ell_{R3j}\ell_{Rj})
(\sin\xi_W W^{+}_\mu + \cos\xi_W W^{'+}_{\mu})  ~+~{\rm h.~c.}, \nonumber \\
{\cal L}_Z &=&\frac{g_L}{ 2} \tan\theta_W (\tan\theta_R + \cot\theta_R) (\sin\xi_Z Z_\mu + \cos\xi_Z Z^\prime_\mu) (\bar \tau_{Ri} V^{\ell*}_{R3i}V^\ell_{R3j}\gamma^\mu \tau_{Rj} - \bar \nu_{R3} \gamma^\mu P_R \nu_{R3} )\nonumber\;,
\label{cccouplep}
\end{eqnarray}
where
\begin{eqnarray}
U^{\ell\dagger} = U^{ \dagger}_L V^\ell_L\;,\;\;
U^{\ell *}_{RLj3} = (U_{RLi3}^{*}V^\ell_{Lij})\;,\;\;
U^\ell_{LRij} = U_{LR3i} V^\ell_{R3j}\;,\;\;
U^\ell_{R3j}=U_{R33} V^\ell_{R3j}\;.
\end{eqnarray}
$U^\ell$ is approximately the PMNS matrix.
From Eqs.~\ref{cccoup} and \ref{cccouplep} we see that a large $g_R/g_L$ will enhance the third generation interactions with $W^\prime$.

In terms of neutrino mass eigenstates, 
\begin{eqnarray}
\bar \nu_{R3} \gamma^\mu \nu_{R3} = - (\bar \nu^m_{Li} U^*_{LRki} + \bar \nu^{mc}_{R_3} U^*_{R33} )\gamma^\mu (U_{LRkj} \nu^m_{Lj} + U_{R33} \nu^{mc}_{R3})\;.
\end{eqnarray}

The new operators in this model that contribute to the rare kaon decay occur at tree level with new FCNC couplings at one-loop with new $Z^\prime$  penguin \cite{He:2004it}. They are 
\begin{eqnarray}
{\cal H}_T&=&-\frac{G_F}{\sqrt{2}}2s^2_W\frac{M_Z^2}{M_{Z^\prime}^2}\cot^2\theta_R V^{d\star}_{Rbs}V^d_{Rbd}\bar{s}\gamma_\mu P_Rd\ \bar\nu_{R3}\gamma^\mu P_R\nu_{R3}\nonumber \\
{\cal H}_L&=&-\frac{G_F}{\sqrt{2}}\frac{\alpha}{\pi}\frac{M_Z^2}{M_{Z^\prime}^2}\cot^2\theta_R V^{\star}_{ts}V_{td} I(\lambda_t,\lambda_H)\bar{s}\gamma_\mu P_Ld\ \bar\nu_{R3}\gamma^\mu P_R\nu_{R3}
\end{eqnarray}
Both contributions couple to the right-handed neutrino so they do not interfere with the SM. In the quark current, only the vector term contributes to a $K\to\pi$  transition so both LH and RH contribute in the same manner to Eq.~\ref{npcase2}.

\section{Models with leptoquarks}

The interest of leptoquarks in kaon decays has been recently revived in connection to the B-anomalies \cite{ Kumar:2016omp,Fajfer:2018bfj}, here we will conduct a model independent analysis as in earlier papers \cite{Davies:1990sc,Davidson:1993qk}. 
The scalar $S$ and vector $V$ leptoquark couplings to SM fermions which include a left-handed neutrino $\nu_L$ are,
\begin{eqnarray}
&&{\cal L}_S = \lambda_{LS_0} \bar q^c_L i \tau_2 \ell_L S_0^\dagger + \lambda_{L\tilde S_{1/2}} \bar d_R \ell_L \tilde S^\dagger_{1/2} + \lambda_{LS_1}\bar q^c_L i\tau_2 \vec \tau \cdot \vec S^\dagger_1 \ell_L + {\rm ~h.~c}.\;,\nonumber\\
&&{\cal L}_V =  \lambda_{L\tilde V_{1/2}} \bar d_R^c \gamma_\mu \ell_L \tilde V^{\dagger \mu}_{1/2} + \lambda_{LV_1}\bar q_L \gamma_\mu \vec \tau\cdot  \vec V^{\dagger \mu}_1 \ell_L + {\rm ~h.~c.}\;,
\end{eqnarray}
where the leptoquark fields and their transformation properties under the SM group are given by
\begin{eqnarray}
&&S^\dagger_0 = S_0^{1/3} : (\bar 3, 1, 1/3)\;,\;\;\tilde S_{1/2}^\dagger = \left ( \tilde S_{1/2}^{-1/3}, \tilde S_{1/2}^{2/3} \right ): ( 3, 2,1/6)\;,\nonumber\\
&&\vec \tau\cdot \vec S_1^{\dagger} = 
\left (
\begin{array}{cc} 
S^{1/3}_1&\sqrt{2} S^{4/3}_1\\ 
\sqrt{2} S^{-2/3}_1&-S^{1/3}_1
\end{array}
\right ): (\bar 3, 3, 1/3)\;;\nonumber\\
&&V_{1/2}^\dagger = \left ( V_{1/2}^{1/3}, V_{1/2}^{4/3} \right ): ( \bar 3, 2, 5/6)\;,\nonumber\\
&&\vec \tau\cdot \vec V_1^{\dagger} = 
\left (
\begin{array}{cc} 
V^{2/3}_1&\sqrt{2} V^{5/3}_1\\ 
\sqrt{2} V^{-1/3}_1&-V^{2/3}_1
\end{array}
\right ): ( 3, 3, 2/3)\;.
\end{eqnarray}

Exchange of these leptoquarks at tree-level generates effective operators of the form $\bar d \Gamma d \bar \nu\Gamma \nu$ that will induce the rare kaon decays. We find with the aid of the identities 
\begin{eqnarray}
\bar q_1P_L\nu_2 \bar \nu_3 P_R q_4 &=& -\frac{1}{2}\bar q_1 \gamma_\mu P_R q_4 \bar \nu_3 \gamma^\mu P_L \nu_2 \nonumber \\
\bar q^c_1\gamma^\mu P_R q^c_2 &=& - \bar q_2 \gamma^\mu P_L q_1\nonumber \\
\bar q_1\gamma_\mu P_L\nu_2 \bar \nu_3 \gamma^\mu P_L  q_4 &=& \bar q_1 \gamma_\mu P_L q_4 \bar \nu_3 \gamma^\mu P_L \nu_2
\end{eqnarray}
an effective four-fermion interaction of the form
\begin{eqnarray}
{\cal L}_{eff} &=& \left( \frac{\lambda^{ij}_{LS_0} \lambda^{\star kl}_{LS_0}}{ 2 m_{S_0}^2}  +
\frac{\lambda^{ij}_{LS_1} \lambda^{\star kl}_{LS_1}}{ 2 m_{S_1}^2} - 2\frac{\lambda^{kj}_{L V_1} \lambda^{\star il}_{LV_1}}{ m_{V_1}^2} 
\right)
\bar d_{Lk}\gamma_\mu d_{Li} \bar \nu_{L_l} \gamma^\mu \nu_{Lj}\ \nonumber \\
&+& \left(-\frac{\lambda^{ij}_{L\tilde S_{1/2}} \lambda^{\star kl}_{L\tilde S_{1/2}}}{ 2 m_{S_{1/2}}^2}
+\frac{\lambda^{kj}_{L V_{1/2}} \lambda^{\star il}_{L V_{1/2}}}{ m_{V_{1/2}}^2}
\right)
\bar d_{Ri}\gamma_\mu d_{R k} \bar \nu_{L_l} \gamma^\mu \nu_{Lj}
\end{eqnarray}
For $K\to \pi \nu\bar\nu$ decays they combine to give
\begin{eqnarray}
{\cal L}_{eff} &=&\sum_{lj} \frac{1}{2}c_{l j}\  \bar s \gamma_\mu d \bar \nu_{L_l} \gamma^\mu \nu_{Lj} +{\rm ~h.~c.} \nonumber \\
c_{l j}&=&\left( \frac{\lambda^{1j}_{LS_0} \lambda^{\star 2l}_{LS_0}}{ 2 m_{S_0}^2}  +
\frac{\lambda^{1j}_{LS_1} \lambda^{\star 2l}_{LS_1}}{ 2 m_{S_1}^2} - 2\frac{\lambda^{2j}_{L V_1} \lambda^{\star 1l}_{LV_1}}{ m_{V_1}^2} -\frac{\lambda^{2j}_{L\tilde S_{1/2}} \lambda^{\star 1l}_{L\tilde S_{1/2}}}{ 2 m_{S_{1/2}}^2}
+\frac{\lambda^{1j}_{L V_{1/2}} \lambda^{\star 2l}_{L V_{1/2}}}{ m_{V_{1/2}}^2}
\right)
\label{lqres}
\end{eqnarray}
Of these leptoquarks all but $S_0$ contribute to processes $d\bar{d}\to\nu\bar\nu$, $u \bar{d}\to \ell^+\nu$, $u\bar{u}\to \ell^+\ell^-$and $d\bar{d}\to \ell^+\ell^-$. This usually means that their effects in the kaon sector are severely constrained by $K_L \to \mu e$ which places their mass in the hundreds of TeV for couplings of electroweak strength and above 1000~TeV for Pati-Salam leptoquarks \cite{Valencia:1994cj}. On the other hand $S_0$ does not contribute to $d\bar{d}\to \ell^+\ell^-$ processes and its effects in the kaon sector are mostly constrained by lepton universality in $\pi_{\ell 2}$ and $K_{\ell 2}$ decays, and as we show here, by $K\to \pi \nu \bar\nu$. Leptoquark models produce both LFC and LFV interactions in general so their contribution to the rare kaon rates are generally of the form
\begin{eqnarray}
&&{\cal B}_{K^+}  =\frac{\tilde\kappa_+}{3}\sum_i\left[\left(\frac{{\rm Im}(V^\star_{ts}V_{td})}{\lambda^5}X_t+ {\rm Im}W_{ii}\right)^2
+ \left(\frac{{\rm Re}(V^\star_{cs}V_{cd})}{\lambda}P_c+\frac{{\rm Re}(V^\star_{ts}V_{td})}{\lambda^5}X_t+ {\rm Re}W_{ii}\right)^2 \right]\nonumber \\
&&+\frac{\tilde\kappa_+}{3}\sum_{i\neq j}|W_{ij}|^2 \nonumber \\
&&{\cal B}_{K_L}=  \frac{\kappa_L}{3}\sum_i \left[\left(\frac{{\rm Im}(V^\star_{ts}V_{td})}{\lambda^5}X_t+ {\rm Im}W_{ii} \right)^2
\right]+  \frac{\kappa_L}{3}\sum_{i\neq j}\left|\frac{(W_{ij}-W^\star_{ji})}{2}\right|^2
\label{ratesLQ}
\end{eqnarray}
The $W_{ij}$ parameters appearing here are versions of the $c_{ij}$ in Eq.~\ref{lqres} but with a different normalisation, $c_{ij} \sim \frac{G_F}{\sqrt{2}}\frac{2\alpha}{\pi s^2_W}V^\star_{ts}V_{td} W_{ij}$. In the main text we only consider the effect of the LFV couplings as the LFC ones fall under the same type of NP as Eq.~\ref{npcase1}.

\section{$\Delta I =3/2$ transitions}

To change the GN relation we construct a $\Delta I =3/2$ operator to mediate the $K\to \pi$ transition. This requires four quark fields, and an example of a dimension eight operator consistent with the symmetries of the SM that accomplishes this is
\begin{eqnarray}
{\cal L}_{NP}^\prime&=&i\frac{g_{NP}}{\Lambda^4}\left(\bar{u}\gamma_\nu P_Rs\ \bar{d}\gamma_\mu P_Ru+\bar{d}\gamma_\mu P_Rs\ (\bar{u}\gamma_\nu P_Ru -\bar d \gamma_\nu P_Rd
)\right) \ g^\prime B^{\mu\nu}  +{\rm h.~c.} 
\label{leff27}
\end{eqnarray}
in which $g_{NP}$ is complex. On dimensional grounds this low energy effective operator is dimension eight and was therefore normalised with $\Lambda^4$.  In general there are two possibilities: the operator may arise from  a dimension eight operator describing physics beyond the SM at the electroweak scale in which case $\Lambda^4=\Lambda_{NP}^4$; or it may arise from a dimension six new physics operator. In the latter case one of the quark currents may occur from a long distance photon, for example, and  the scale suppression could be smaller, $\Lambda^4\sim\Lambda_{NP}^2\Lambda_{QCD}^2$ as in
\begin{eqnarray}
{\cal L}_{NP}\sim i \bar d\gamma_\mu D_\nu P_Rs B^{\mu\nu} + {\rm ~h.~c.}
\end{eqnarray}
A possible bosonisation for the four-quark operator in Eq.~\ref{leff27} of the current-current form is, $R^\mu_{21}R^\nu_{13}+R^\mu_{23}(R^\nu_{11}-R^\nu_{22})$, where $R_\mu=if_\pi^2U^\dagger D_\mu U$ and $U=\exp(2i\phi/f_\pi)$ with $\phi$ the pseudoscalar meson octet as usual in chiral perturbation theory \cite{Kambor:1989tz}. This allows us to write Eq.~\ref{leff27}  as, \footnote{We are not interested in finding the most general representation of the operator in chiral perturbation theory, but only in giving an example of what the matrix element might look like.}
\begin{eqnarray}
{\cal L}_{NP}&=& 2 \frac{g_{NP}s_W^2}{\Lambda^4}\ f_\pi f_K\left( \sqrt{2} \partial_\mu K^0\partial_\nu\pi^0+ \partial_\mu K^+\partial_\nu\pi^-  \right)\ \frac{g}{c_W}Z^{\mu\nu}
+{\rm h.~c.}
\end{eqnarray}
which then leads to matrix elements
\begin{eqnarray}
{\cal M}(K^+ \to \pi^+ \nu\bar\nu)_{NP}&=&\frac{G_F}{\sqrt{2}}\frac{2\sqrt{2}s_W^2 f_\pi f_K}{\Lambda^4}\ g_{NP}\ m_{\nu\nu}^2\ 2p_K^\mu\sum_\ell\bar\nu_\ell\gamma_\mu P_L \nu_\ell, 
\nonumber \\
{\cal M}(K_L \to \pi^0 \nu\bar\nu)_{NP}&=&\frac{G_F}{\sqrt{2}}\frac{4\sqrt{2} s_W^2f_\pi f_K}{\Lambda^4}\ {\rm Im}(g_{NP})\ m_{\nu\nu}^2\ 2p_K^\mu\sum_\ell\bar\nu_\ell\gamma_\mu P_L \nu_\ell.
\label{newmatel}
\end{eqnarray}
Compared to the matrix elements of the operators discussed in previous sections there is an additional $m_{\nu\nu}^2$ term in Eq.~\ref{newmatel}. This modifies the rates by the factor
\begin{eqnarray}
r^2_{ps}m_K^4&\equiv&\frac{\int d\Phi_3 \left|m_{\nu\nu}^2 p_K^\mu\bar\nu\gamma_\mu P_L \nu\right|^2}{\int d\Phi_3 \left| p_K^\mu\bar\nu\gamma_\mu P_L \nu\right|^2} \approx (0.171m_K^2)^2 
\end{eqnarray}
so that  $r_{ps}=0.171$ for $K^+$ decay, and $r_{ps}=0.176$  for $K_L$ decays.

It is possible to write an analogue of Eq.~\ref{leff27} using left-handed quark fields at the expense of higher dimensionality. An example being,
\begin{eqnarray}
{\cal L}_{NP}^\prime&\sim& \frac{g_{NP}}{\Lambda^6}\left(\bar{q}_2\gamma_\mu \tau^IP_Lq_1 \ \bar{q}_1\gamma_\nu \tau^IP_Lq_1\ \Phi^\dagger \Phi -3  \bar{q}_2\gamma_\mu P_Lq_1 \ \bar{q}_1\gamma_\nu \tau^IP_Lq_1\ \Phi^\dagger  \tau^I \Phi \right) \ g^\prime B^{\mu\nu}  +{\rm h.~c.} 
\label{leff27lf}
\end{eqnarray}
where ${1,2}$ are generation indices, $\Phi$ is the SM scalar doublet, and additional flavour changing operators involving charm are also produced.

In principle one could start with an operator at the electroweak scale with a flavour structure  such that it contributes only to the neutral kaon decay. This would be a mixture of $\Delta_I=1/2$ and $\Delta I =3/2$ operators and the two components would evolve differently under QCD running resulting in a different flavour structure at the hadronic scale.

\end{document}